\documentclass[a4paper,11pt]{article}
\usepackage{pos}
\usepackage{siunitx}
\usepackage{subcaption}
\DeclareSIUnit\year{yr}
\DeclareSIUnit\parsec{pc}

\title{DUNE's low energy physics searches}

\author*[a]{Sergio Manthey Corchado}

\onbehalf{for the DUNE collaboration}

\affiliation[a]{CIEMAT,\\
  Avenida Complutense 40, Madrid, Spain}

\emailAdd{sergio.manthey@ciemat.es}

\abstract{
The Deep Underground Neutrino Experiment (DUNE) is a long--baseline neutrino experiment that will precisely measure neutrino oscillation parameters, observe astrophysical neutrinos, and search for processes beyond the Standard Model such as nucleon decays, heavy neutral leptons, and dark matter. DUNE will build four Liquid Argon Time Projection Chambers (LAr--TPC), as far detectors, with a total mass of $\sim\SI{70}{\kilo\tonne}$ LAr located at Sanford Underground Research Facility (SURF), \SI{1.5}{\kilo\metre} underground. A near--site complex, hosting different detectors, will measure the neutrino flux from an accelerated particle beam (\SI{1.2}{\mega\watt}) produced at the Long Baseline Neutrino Facility (LBNF) at Fermilab, \SI{1300}{\kilo\metre} away from SURF. Additionally, the few--MeV low energy regime is of particular interest for detecting the burst of neutrinos from a galactic core-collapse supernova and solar neutrinos. DUNE will measure oscillation parameters from the solar neutrino flux analysis with much higher precision than previous experiments and discover the yet--unobserved hep flux component. 
}

\FullConference{%
  17-24 July 2024\\
  Prague Congress Centre, Prague, Czech Republic\\
  International Conference on High Energy Physics - ICHEP 2024
}

\begin{document}
\maketitle

\section{Low Energy Interactions}\label{sec:low_energy_searches}
the dominant neutrino interaction channel in LAr is the charged-current (CC) absorption of an electron neutrino ($\nu_e$) on $^{40}\text{Ar}$, with a Q--value of \SI{1.5}{\mega\eV} and cross--section ranging from \qtyrange{1e-43}{1e-38}{\square\centi\meter}. The observable topology is a short electron track surrounded by deexcitation gammas ($\gamma$), which appear as TPC blips, up to \SI{1}{\meter} apart from the de--excitation of the resulting $^{40}\text{K}^\text{*}$ nucleus. Less frequent, but equally important, is the elastic scattering (ES) of neutrinos on electrons, characterized by tracks and the absence of gammas, that conserve some directional information of the incoming particle. This channel is fundamental to performing pointing studies and can provide a flavor--independent flux measurement. According to current simulation standards, DUNE's reconstruction threshold corresponds to $\sim\SI{1}{\mega\eV}$ (but could potentially be lower), with an energy resolution between 10 and 20\% in the few \unit{\mega\eV} range, allowing the experiment to access a full range of low energy searches. In the following, we focus on two prominent examples.


\subsection{Solar Neutrinos}\label{sec:solar_neutrinos}

\begin{figure*}[!h]
\centering
\begin{subfigure}[t]{0.44\textwidth}
    \centering
    \includegraphics[width=1\textwidth]{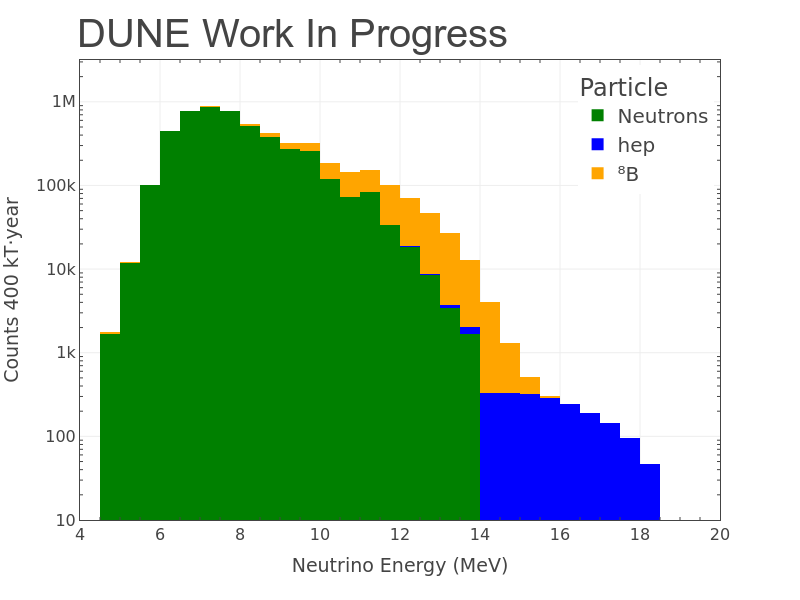}
    \caption{Reconstructed solar neutrino energy spectrum.}
    \label{fig:solar_flux}
\end{subfigure}\quad
\begin{subfigure}[t]{0.48\textwidth}
    \centering
    \includegraphics[width=1\textwidth]{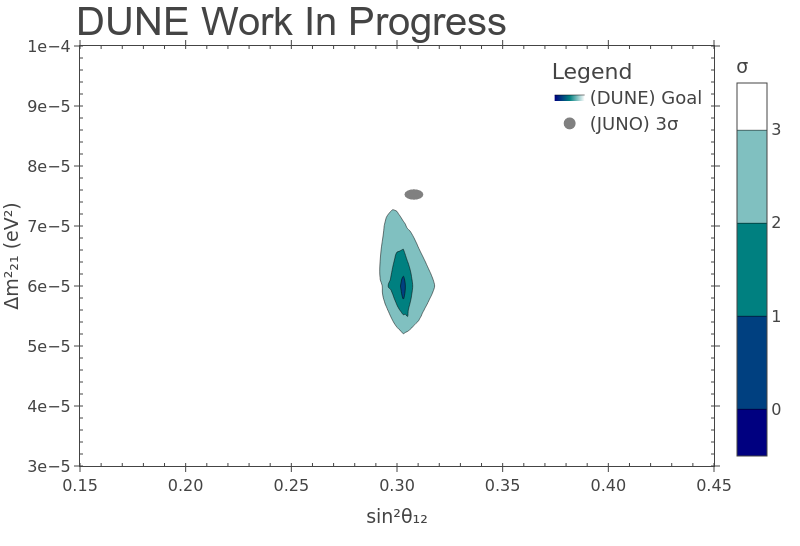}
    \caption{DUNE's sensitivity contours to $\Delta \text{m}_{21}^2$ and $\sin{^{2}\theta}_{12}$.}
    \label{fig:solar_contours}
\end{subfigure}
\end{figure*}

Our Sun produces a continuous flux of $\nu_e$ as a by--product of fusion reactions in its core. DUNE has the potential to characterize and study the oscillation of the two most energetic solar neutrino chains (see Figure \ref{fig:solar_flux}) with reconstructed energies centered at \SI{10}{\mega\eV} and an expected interaction rate of $\sim\SI{10}{\milli\hertz}$. The measurement of \textbf{$^{8}$B} $\nu_\text{e}$ flux is motivated by the existing tension (recently reduced by the latest Super--Kamiokande results) between the preferred $\Delta \text{m}_{21}^{2}$ value measured by reactor- and solar--neutrino experiments \cite{capozzi2019dune}. A background model accounting for intrinsic LAr isotopes and contaminants, detector components, and even external radiation from SURF's cavern has been implemented. Within this realistic model, neutron captures from the cavern are expected to be the measurement's main background due to a similar topology to CC solar $\nu_\text{e}$ events and a deep penetration length avoiding rejection by fiducialization. The highest and irreducible capture energy (released as a $\gamma$ cascade from the nucleus) corresponds to $\sim\SI{9}{\mega\eV}$ (n + $^{36}\text{Ar}\rightarrow \gamma^\text{s}$) which could be reconstructed up to \SI{14}{\mega\eV} fake neutrino signals. Assuming DUNE's exposure of \SI{400}{\kilo\tonne\year}, current studies estimate a Signal--to--Background of \num{1.19} with $>400$k events above \SI{10}{\mega\eV} ($\sim 2$k events above \SI{14}{\mega\eV} in the \textbf{hep} region) and predicts sensitivities to $\Delta \text{m}_{21}^{2}$ with $>3\sigma$ separation to projected results from the JUNO experiment (see Figure \ref{fig:solar_contours}). Studies including reconstructing the drift--coordinate by matching the charge and light signals are ongoing. 

\subsection{Supernova Neutrino Detection and Pointing}\label{sec:supernova_neutrinos}

\begin{figure*}[!h]
\centering
\begin{subfigure}[t]{0.52\textwidth}
    \centering
    \includegraphics[width=1\textwidth]{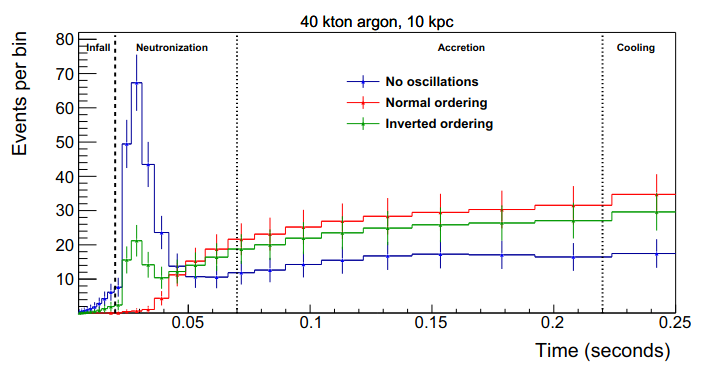}
    \caption{Time spectrum of a neutrino burst from a core-collapse supernova up to \SI{10}{\kilo\parsec}.}
    \label{fig:supernova_time_spectra}
\end{subfigure}
\quad
\begin{subfigure}[t]{0.42\textwidth}
    \centering
    \includegraphics[width=1\textwidth]{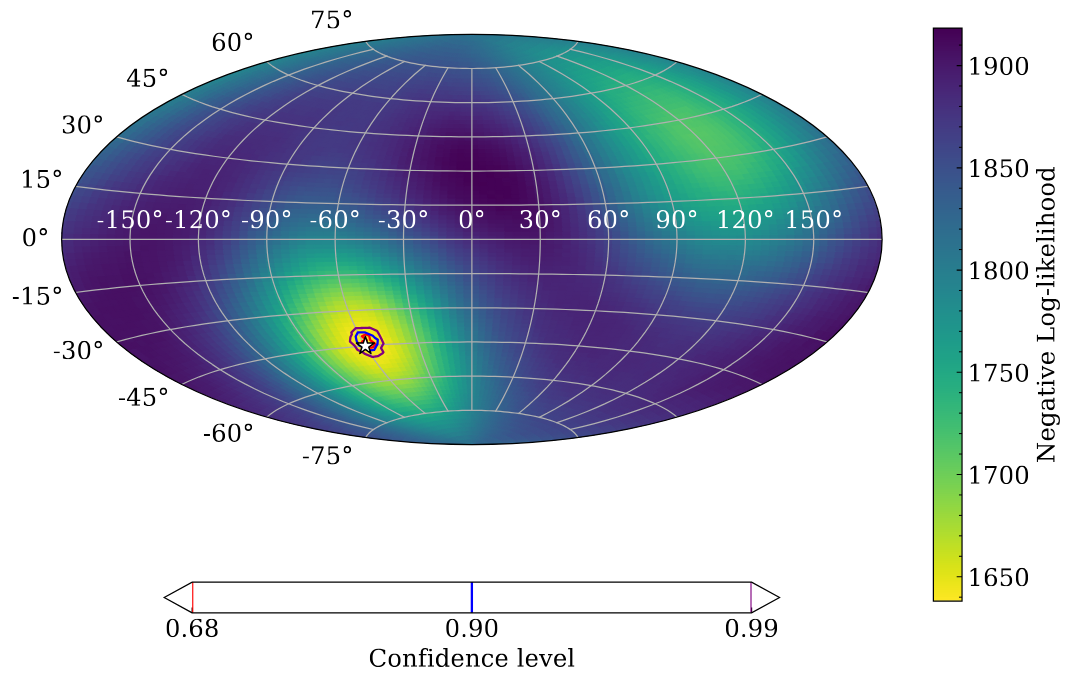}
    \caption{Reconstructed pointing example from a simulated core--collapse supernova burst}
    \label{fig:supernova_pointing}
\end{subfigure}
\end{figure*}

A supernova releases an intense source of neutrinos of all flavors (99\% of the released energy), which are ideal astrophysical messengers due to their prompt and direct travel upon release. DUNE can probe the inner evolution of the core-collapse mechanism by studying the neutrino's arrival time and energy spectra \cite{abi2021supernova}. Within DUNE, three qualitative stages of the collapse can be distinguished (see Figure \ref{fig:supernova_time_spectra}). Preliminary studies show good sensitivity to the entire Milky Way, and possibly beyond, with thousands of interacting neutrinos (model--dependent). By measuring the spectral profiles and non--thermal features, DUNE will be sensitive to a broad range of supernova and neutrino physics phenomena, including neutrino mass ordering and collective effects.\\

By reconstructing the ES neutrino events during a supernova burst, its direction can be inferred. To minimize the head-tail ambiguities in track reconstruction, a new algorithm called "brems--flipping" has been developed \cite{abud2024supernova}. So far, pointing resolution has been estimated under different reconstruction scenarios. \textbf{Perfect:} (ideal classification of ES events) resolution is 3.4 degrees (6.6 degrees) with 68\% coverage for supernovae at a distance of 10 kpc and an effective fiducial volume of \SI{40}{\kilo\tonne} (\SI{10}{\kilo\tonne}). \textbf{Optimistic} (incorrect classification of 4\% $\nu_e$ CC events as ES): resolution is 4.3 degrees (8.7 degrees). Accurate and fast determination is critical for multi-messenger astronomy, and efforts are ongoing to make pointing with DUNE reliable and efficient. Due to the presence of mainly CC events, machine--learning techniques will need to be implemented to reject non--ES tracks from contributing to the algorithm.

\bibliography{bibliography} 
\bibliographystyle{ieeetr} 

\end{document}